\documentclass[reprint,amsmath,amssymb,aps,prl,showpacs]{revtex4-1}
\usepackage[utf8]{inputenc}
\usepackage{graphicx}
\usepackage{dcolumn}
\usepackage{bm}
\usepackage[breaklinks]{hyperref}
\usepackage{bbold}
\usepackage{color}
\usepackage[normalem]{ulem}
\usepackage{cancel}
\renewcommand*{\HyperDestNameFilter}[1]{\jobname-#1}

\renewcommand{\vec}[1]{\mathbf{#1}}
\newcommand{\vn}[1]{{\vec{#1}}}
\newcommand{\vht}[1]{{\boldsymbol{#1}}}
\newcommand{\hatn}{{\hat{\vec n}}}
\newcommand{\vecr}{\vec R}

\newcommand{\lambdaso}{\lambda_{\text{so}}}
\newcommand{\bex}{\vec b^{\text{ex}}}
\newcommand{\gso}{\vec g^{\text{so}}}

\begin{document}

\preprint{APS/123-QED}

\title{Phase-Space Berry Phases in Chiral Magnets: Dzyaloshinskii-Moriya Interaction and the Charge of Skyrmions}

\author{Frank  Freimuth$^{1}$}
\email{f.freimuth@fz-juelich.de}
\author{Robert Bamler$^{2}$}
\author{Yuriy Mokrousov$^{1}$}
\author{Achim Rosch$^{2}$}
\affiliation{$^1$Peter Gr\"unberg Institut and Institute for Advanced Simulation,
Forschungszentrum J\"ulich and JARA, 52425 J\"ulich, Germany}
\affiliation{$^2$Institute for Theoretical Physics, Universit\"at zu K\"oln, D-50937 K\"oln, Germany}

\date{\today}

\begin{abstract}
The semiclassical motion of electrons in phase space, $\vec x=(\vecr,\vec k)$, is influenced by Berry phases described by a 6-component vector potential, $\vec A=(\vec A^{\rm R}, \vec A^{\rm k})$.
In chiral magnets Dzyaloshinskii-Moriya (DM) interactions induce slowly varying magnetic textures (helices and skyrmion lattices) for which
all components of $\vec A$ are important inducing effectively a curvature in mixed position and momentum space. We show that for smooth textures and weak spin-orbit coupling phase space Berry curvatures determine  the DM  interactions and give important contributions to the charge.
Using {\em ab initio} methods we calculate the strength of DM interactions in MnSi in good agreement with experiment and estimate the charge of skyrmions.
\end{abstract}

\maketitle

In chiral magnets without inversion symmetry, spin-orbit interaction (SOI) effects described
by Dzyaloshinskii-Moriya (DM) interactions~\cite{dmi_dzyalo,dmi_moriya} induce the formation of magnetic textures. For smooth textures and cubic systems like MnSi, the leading DM contribution to the free energy density is given by the term $D \hatn \cdot(\nabla \times \hatn)$, where $\hatn$ is the direction of the magnetization. This term describes that energy can be gained when the magnetic structure twists.
In small magnetic fields these interactions (in combination with thermal fluctuations) can stabilize lattices of topologically quantized magnetic whirls, so called skyrmions \cite{muhlbauer_skyrmion_2009,yu_real-space_2010}. Skyrmions
couple due to their topological winding extremely  efficiently to electric currents resulting in ultralow critical currents for the motion of skyrmions \cite{schulz_emergent_2012,everschor_rotating_2012,yu_skyrmion_2012}.

%

In this letter we argue that Berry phases in phase space provide not only a natural framework to understand the physical properties of skyrmions and other magnetic textures but also generate DM interactions and act therefore as 
the main driving force inducing magnetic textures in chiral magnets. We focus on 
Berry curvatures in 
mixed position and momentum space which lead both to  DM interactions and also to an electric charge of skyrmions. Thereby we naturally link skyrmions in chiral magnets to skyrmions in quantum Hall systems with filling close to $\nu=1$, which are characterized by a quantized {electric} charge \cite{barrett_optically_1995,fertig_charged_1994,sondhi_skyrmions_1993,lee_boson-vortex-skyrmion_1990,brey_skyrme_1995}.

Berry phases are quantum mechanical phases picked up by a quantum system when the wave function changes adiabatically \cite{berry_quantal_1984,sundaram_wave-packet_1999,xiao_berry_2010}.  They can strongly affect the semiclassical motion of electrons. For each electronic band $n$, the effects of smoothly varying magnetic textures can efficiently be described by a six component vector potential, $\vec A_n=(\vec A_n^{\rm R}, \vec A_n^{\rm k})$,
with 
\begin{eqnarray}\label{berry-connection}
A_{n,j}(\vec x)= \langle \vec x, n | i\frac{\partial}{\partial x_j}| \vec x, n \rangle,\quad  j=1,...,6
\end{eqnarray}
where $\vec x=(\vecr,\vec k)$ is the position in phase space and $ |\vec x, n \rangle=  | \hatn (\vec R),\vec k,n \rangle$ is the Bloch function, which depends not only on lattice momentum $\hbar\vec k$, but also on the orientation $\hat {\vec n}(\vecr)$ of the magnetization. Here we use the letter $\vecr$ to denote smooth variations on length scales much larger than the lattice spacing.

Two aspects of Berry phase physics have been well studied in the context of chiral magnets.
First, Berry phases in momentum space, described by the $\vec k$ dependence of $\vec A_n^{\rm k}(\vec x)$, give rise to the anomalous Hall effect~\cite{nagaosa_anomalous_2010}, which dominates the Hall response for a wide range of temperatures and fields in materials like MnSi~\cite{lee_hidden_2007}. Powerful {\em ab initio}  methods have been developed to calculate the anomalous Hall effect quantitatively~\cite{ahe_bcc_iron,ahe_wannier_interpolation,ahe_kkr_cpa,ahe_scattering_independent,ahe_amr_tblmto}. Second, real-space Berry phases give rise to the so-called topological Hall effect.
For weak SOI, each skyrmion contributes due to their topology one flux quantum of an emergent magnetic flux \cite{neubauer_topological_2009},
arising from the effective magnetic field
$B^{\rm R}_{n,i}=\frac{\hbar}{e} \epsilon_{ijk} \partial_{R_j} A^{\rm R}_{n,k} \approx \pm \frac{\hbar}{4e} \epsilon_{ijk}\hat{\vec n} \cdot  ( \partial_{R_j}  \hat{\vec n} \times \partial_{R_k} \hat{\vec n})$ with positive (negative) sign for majority (minority) band $n$, respectively. This real-space emergent magnetic field acts similar 
to the ``real'' magnetic field and has been observed in MnSi \cite{neubauer_topological_2009,muhlbauer_skyrmion_2009} and other materials (see, e.g., \cite{kanazawa_large_2011,huang_extended_2012}) as an extra contribution to the Hall signal. The same effect is also responsible for the efficient coupling of
skyrmions to electric currents \cite{schulz_emergent_2012,everschor_rotating_2012,yu_skyrmion_2012}.

Much less studied are systems with Berry phases in phase space, where 
the $\vecr$ dependence of $\vec A_n^{\rm k}(\vec x)$ and the $\vec k$ dependence of $\vec A_n^{\rm R}(\vec x)$ become important. It has been argued that such a situation arises in smoothly deformed crystals \cite{sundaram_wave-packet_1999} or
in the presence of spatially varying external magnetic fields. Also in antiferromagnets with slowly varying spin texture 
the mixed Berry phase was
suggested to crucially influence the adiabatic dynamics of electrons \cite{cheng_electron_2012}. Some of us \cite{mothedmisot} have recently pointed out that DM interactions arise from certain Berry phases. Here we will provide a purely semiclassical derivation of the Berry phase contribution to the DM interaction showing that DM interactions can be viewed as a phase-space Berry phase effect. By the same mechanism, magnetic skyrmions also obtain a charge.
In general, chiral magnets and their magnetic phases turn out to be ideal model systems to study phase-space Berry phases due to their smoothly varying magnetic textures driven by DM interactions.

As has been shown by Niu {\it et al.}~\cite{sundaram_wave-packet_1999,xiao_berry_2005}, phase-space Berry phases effectively lead to a curvature of phase space  described by the antisymmetric $6 \times 6$ Berry-curvature tensor \cite{xiao_berry_2005}
\begin{equation}\label{berry-curvature}
\Omega_{n,ij} = \frac{\partial A_{n,j}}{\partial x_i}-\frac{\partial A_{n,i}}{\partial x_j}= \begin{pmatrix}
		\Omega_n^{\rm RR}	& \Omega_n^{\rm Rk} \\
		\Omega_n^{\rm kR}	& \Omega_n^{\rm kk}
	\end{pmatrix}_{ij}.
\end{equation}
Here, $\Omega^{\rm RR}_{n,ij}= \frac{e}{\hbar} \epsilon_{ijk} B^{\rm R}_{n,k}$ describes real-space Berry phases while
$\Omega^{\rm kk}_{n,ij}$ encodes the momentum space Berry phases also discussed above.  The $3\times 3$ matrix $\Omega^{\rm Rk}_{n,ij} dR_i dk_j$ is the Berry phase which is picked
up when an electron moves along a loop in the phase-space plane spanned by the coordinates $R_i$ and $k_j$. 

The Berry phases influence the semiclassical description of the system in three points.
First, the combination of smooth variations in both position and momentum space leads to a shift of the semiclassical energy levels \cite{sundaram_wave-packet_1999}, 
$\epsilon_n(\vec x) = \epsilon_n^{(0)}(\vec x) + \delta\epsilon_n(\vec x)$ where $\epsilon_n^{(0)}(\vec x)=\langle \vec x,n|H(\vec x)|\vec x,n\rangle$ and
\begin{equation}\label{semiclass-eshift}
	\delta\epsilon_n(\vec x) = -\text{Im}\left[ \frac{\partial \langle \vec x,n|}{\partial R_i} (\epsilon_n^{(0)}(\vec x) - H(\vec x)) \frac{\partial |\vec x,n\rangle}{\partial k_i} \right].
\end{equation}
Second, the Berry phases modify the semiclassical equations of motion \cite{sundaram_wave-packet_1999}, which read $(\Omega_n-J)\dot{\vec x}=\frac{\partial \epsilon_n}{\partial \vec x}$ where $J= \begin{pmatrix}
	0 & \mathbb 1 \\
	-\mathbb 1 & 0
\end{pmatrix}$.
Third, the curvature of phase space leads in 
semiclassical approximation to a modified density of states in phase space \cite{morrison_hamiltonian_1998,xiao_berry_2005},
\begin{alignat}{1}
	W_n(\vec x) &= \sqrt{\text{det}(\Omega_n\!-\!J)} \nonumber \\
	&= \frac{\epsilon_{ijklrs}}{48}	(\Omega_n\!-\!J)_{ij}(\Omega_n\!-\!J)_{kl}(\Omega_n\!-\!J)_{rs}\;. \label{semiclass-dos}
\end{alignat}
Only for this modified density of states the Liouville theorem holds.
A derivation of Eq.~(\ref{semiclass-dos}) is given in the supplement \cite{supplement}.

Within density functional theory one can describe the ground state of a ferromagnetic many particle system 
with magnetization parallel to the unit vector $\hatn$ by an effective single-particle Kohn-Sham Hamiltonian
\begin{equation}\label{eq_local_hamiltonian}
H_{\hat{ \vec n}}=\frac{{\vec p}^2}{2m}+V(\vn{r})- \vec M \cdot \vec{B}(\vec r)
-\frac{1}{2mc^2}\vec{M}
\cdot
\left(
\vn{E}(\vn{r})\times \vn{p}
\right),
\end{equation}
parametrized by an effective potential $V(\vn{r})$, an exchange field $\vec{B}(\vec r)$ and an electric field $\vn{E}(\vn{r})$. 
We use the letter $\vec r$ for variations on the atomic length scale. To obtain such a ferromagnetic state in a chiral magnet, one has 
to apply a small external field $\bf B^{\rm ext}$ in the direction of $\hatn$ (implicitly included in  $\vec{B}$), see below. In the absence of SOI, $V$ and $\vn{E}$ are independent of $\hatn$, while
$\vec{B} \| \hatn$.

Starting from the eigenstates  $|\hatn,\vec k,n\rangle$ of the {\em uniform} Hamiltonian (\ref{eq_local_hamiltonian}) for fixed $\hatn$, we can obtain the change of the free energy density, $\delta F^{(1)}(\vn{R})$, to leading order in an adiabatic approximation. We assume that $\hatn(\vec R)$ slowly varies in space and use Eq. (\ref{semiclass-eshift}) and (\ref{semiclass-dos}) with $|\vec x,n\rangle=|\hatn(\vec R),\vec k,n\rangle$ to obtain
\begin{equation}\label{semiclass-deltaf}
\begin{aligned}
\delta F^{(1)}(\vn{R})&=
\sum_{n}
\int\frac{d^3 k}{(2\pi)^{3}}
[
f_{\vn{k}n}\delta\epsilon_{n}(\vn{x})\\
&+
\frac{1}{\beta}
\ln(1+e^{-\beta(\epsilon_{\vn{k}n}-\mu)})\Omega^{\rm Rk}_{nii}(\vn{x})
].
\end{aligned}
\end{equation}
Note that to leading order all contributions arise from mixed position and momentum space Berry curvatures (see below). They contribute only when both inversion symmetry is broken and SOI is present.

To calculate $\delta\epsilon_{n}(\vn{x})$ and the Berry curvature $\Omega^{\rm Rk}_{nii}(\vn{x})$ directly, we use that
the change of an eigenstate $| n  \rangle$ upon changing a parameter $\lambda$ of $H$ is given by $\partial_{\lambda} |n \rangle=\sum_{m\neq n} \frac{|m \rangle}{ E_n-E_m} \left\langle m| \frac{\partial H}{\partial \lambda} | n \right \rangle$ and therefore the Berry curvature reads
\begin{equation}\label{abinitio-Omega}
%
\Omega_{n,ij}=- 2 \left[\sum_{m \neq n}   \Im \frac{\left\langle \vec k n| \frac{\partial H}{\partial x_i} | \vec k m \right \rangle \left\langle \vec k m| \frac{\partial H}{\partial x_j} | \vec k n \right \rangle}{(\epsilon_{\vec k n}-\epsilon_{\vec k m})^2}\right].
\end{equation}
The derivative with respect to crystal momentum is identified with the velocity,
$\vec v$, while the derivative with respect to position arises from the $\vecr$-dependence of $\hatn$
\begin{equation}\label{derivatives}
\frac{1}{\hbar}\frac{\partial H}{\partial  k_i}=v_i, \quad
\frac{\partial H}{\partial  R_i}=
\frac{\partial H}{\partial \hat{\vec n}}\cdot\frac{\partial \hat{\vec n}}{\partial R_i}=
\vec T(\vec r) \cdot
\left(
\hat{\vec n}
\times
\frac{\partial \hat{\vec n}}{\partial  R_i}
\right)
\end{equation}
where $\vec T(\vec r)=\hatn \times \frac{\partial H}{\partial \hatn}$ 
is the torque operator.
Thus, we arrive at
\begin{equation}\label{eq_firstorder_free}
\delta F^{(1)}(\vn{R})=D_{ij}\hat{\vn{e}}_{i}
\cdot
\left(
\hat{\vn{n}}\times\frac{\partial \hat{\vn{n}}}{\partial R_{j}}
\right),
\end{equation}
with 
\begin{eqnarray}
\label{eq_dmi_finite_temperature}
D_{ij}
&=&\sum_{n}
\int\frac{d^3 k}{(2\pi)^3}
f_{\vn{k}n}A_{\vn{k}nij}
+\frac{\ln [1+e^{-\beta(\epsilon_{\vn{k}n}-\mu)}]
B_{\vn{k}nij}}{\beta}\nonumber \\
\label{eq_akn_kubo}
A_{\vn{k}nij}&=&\hbar\sum_{m\neq n}\Im
\left[
\frac{
\langle {\vn{k}n}  |T_{i}| {\vn{k}m}  \rangle
\langle {\vn{k}m}  |v_{j}(\vn{k})| {\vn{k}n}  \rangle
}
{
\epsilon_{\vn{k}m}-\epsilon_{\vn{k}n}
}
\right]\\
B_{\vn{k}nij}
&=&-2\hbar\sum_{m\neq n}\Im
\left[
\frac{
\langle {\vn{k}n}  |T_{i}| {\vn{k}m}  \rangle
\langle {\vn{k}m}  |v_{j}(\vn{k})| {\vn{k}n}  \rangle
}
{
(\epsilon_{\vn{k}m}-\epsilon_{\vn{k}n})^2
}
\right].\nonumber
\end{eqnarray}
where $A_{\vn{k}nij}$ describes the Berry energy  (\ref{semiclass-eshift}), $\delta \epsilon_n=A_{\vn{k}ni'i}
(
\hat{\vn{n}}\times\frac{\partial \hat{\vn{n}}}{\partial R_{i}}
)_{i'}$, and $B_{\vn{k}nij}$ the Berry curvature $\Omega^{\rm Rk}_{n,ij}=B_{\vn{k}ni'j} 
(
\hat{\vn{n}}\times\frac{\partial \hat{\vn{n}}}{\partial R_{i}}
)_{i'}$, respectively.

Eq.~\eqref{eq_firstorder_free} can directly be identified with the DM interaction~\cite{dmi_dzyalo,dmi_moriya} in the continuum limit. We have therefore shown that for smooth textures and weak
SOI the DM interaction arises from mixed momentum and position space Berry curvatures and the corresponding energy shifts obtained from the
Kohn Sham Hamiltonian.
Our semiclassical derivation yields the same expression for $D_{ij}$ as obtained from quantum
mechanical perturbation theory~\cite{mothedmisot}. 

At $T=0$, to linear order in $\bm \nabla \hatn$ and for weak SOI the formula (\ref{eq_firstorder_free}) is exact even for a fully interacting quantum system (provided the exact Kohn Sham Hamiltonian is used in Eq.~(\ref{eq_local_hamiltonian})). To calculate changes of the ground state energy to linear order, changes of $H$ due to $\bm \nabla \hatn$ can be neglected (a manifestation of the magnetic force theorem~\cite{force_theorem}). Therefore the only remaining source of errors is the external magnetic field $\bm B^{\rm ext}$ needed to stabilize the ferromagnetic solution underlying Eq. (\ref{eq_local_hamiltonian}). As this field can be chosen to be weak (second order in SOI strength), it does not affect the value of the DM interaction to leading order in SOI.


Besides DM interactions also the current-induced spin-orbit torque relies on broken inversion symmetry. Recently, it has
been shown that the 
intrinsic contribution to this torque is related to the 
Berry curvature $B_{\vn{k}nij}$~\cite{ibcsoit}.  

The change of charge density to first order in the 
gradients of the magnetization also arises from both the 
change of the density of states and the energy levels from phase space curvatures. It is given by
\begin{equation}\label{rho1}
\delta \rho^{(1)}(\vn{R})=
e \sum_{n}
\int\frac{d^3 k}{(2\pi)^{3}}
\frac{\partial f_{\vn{k}n}}{\partial \epsilon} \delta\epsilon_{n}(\vn{x})- f_{\vn{k}n}\,\Omega^{\rm Rk}_{nii}(\vn{x})
\end{equation}
where $e=-|e|$ is the electron charge. We obtain
\begin{eqnarray}
\delta\rho^{(1)}(\vn{R})&=&e\, G_{ij}\hat{\vn{e}}_{i}
\cdot
\left(
\hat{\vn{n}}\times\frac{\partial \hat{\vn{n}}}{\partial R_{j}}
\right),
\\G_{ij}
&=&
\int\frac{d^3 k}{(2\pi)^3}
\sum_{n}\left[\frac{\partial f_{\vn{k}n}}{\partial \epsilon_{\vn{k}n}}A_{\vn{k}nij}
-f_{\vn{k}n}
B_{\vn{k}nij}
\right].\nonumber
\end{eqnarray}

In metals
extra charges are screened on the length scale set by the Thomas-Fermi 
screening length $\lambda_{TF}$, resulting in a strongly suppressed charge density
$\rho_{\rm tot}\approx - \lambda_{TF}^2 \nabla^2 (\delta \rho^{(1)})$.
Since the metal screens extra charge by changing the occupation of the states on the Fermi surface, while the Fermi sea
participates in the formation of $\delta \rho^{(1)}$, the calculation of 
the unscreened $\delta \rho^{(1)}$ is interesting to understand
how many electrons are
energetically redistributed between Fermi surface and Fermi sea due to the
phase space Berry phases. 

The results of our semiclassical derivation can be reproduced by a gradient expansion of the Green's function similar to the technique used by Yang {\it et al.}~in \cite{yang_skyrmion_2011} for insulators, see supplement \cite{supplement}.
While to leading order in $\bm \nabla \hatn$ the semiclassical formulas Eqs.~\eqref{semiclass-eshift} and \eqref{semiclass-dos} are reproduced, higher orders give for metals rise to additional contributions to the density of states that are not captured by the higher order terms of Eq.~(\ref{semiclass-dos}) as we have checked explicitly.

In insulators the situation is different.
Neither energy shifts, Eq.~(\ref{semiclass-eshift}), nor the term linear in $\Omega_n$ in Eq.~(\ref{semiclass-dos}) contribute to the total charge of a single skyrmion.
The integral  $\int \frac{ dR_x d k_x}{2 \pi} \Omega^{\rm Rk}_{n,xx}$, for example, has to be quantized (first Chern number).
As it evaluates to $0$ for $y \to \infty$, it  vanishes everywhere.
All contributions to the charge arise from higher order terms in the gradient expansion.
As the charge has to be a topological invariant in an insulator, it can be calculated from an adiabatically deformed  band structure where all occupied bands are completely flat and degenerate.
In this limit one can use standard arguments \cite{qi_topological_2008} to show that the 
total accumulated charge due to smooth variations in phase space, e.g., in two space dimensions, is given by the second Chern number
\begin{equation}
	\delta Q \equiv \int\!\!d^2R\; \delta\rho^{(2)} =- e\int\!\! \frac{ d^2 R \, d^2 k}{(2 \pi)^2}\frac{\epsilon_{i j k l}}{8} \text{Tr}\left[ \Omega_{i j} \Omega_{k l}\right]\label{insulator}
\end{equation}
where $\Omega$ is a matrix in the space of occupied bands. Therefore either Abelian or non-Abelian winding numbers can occur.
In cases when all non-Abelian winding numbers vanish, the right-hand side of Eq.~(\ref{insulator}) coincides with the integral over the term quadratic in $\Omega_n$ on the right-hand side of Eq~(\ref{semiclass-dos}).
For Abelian situations, $\delta Q$ can be expressed as a product of two simple real-space and momentum-space winding numbers (i.e., two first Chern numbers), $\delta Q = -\sigma_{xy} \Phi_0$, where $\sigma_{xy}=\frac{e^2}{\hbar} \int \frac{d^2 k}{(2 \pi)^2} \Omega^{\rm kk}_{xy}$ is the quantized Hall conductivity and 
$\Phi_0=\frac{\hbar}{e} \int d^2 R \Omega^{\rm RR}_{xy}$ the quantized total flux arising from the 
real-space Berry phases.  This can be shown by rewriting Eq.~(\ref{insulator}) as a surface integral and using, for example, that
for $\vec R \to \infty$ both $\Omega^{\rm RR}$ and $\Omega^{\rm Rk}$ vanish, see supplement \cite{supplement} for details.

In order to investigate first qualitatively how the accumulated charge in metals depends on the strength of SOI, we consider the simple two-dimensional
two-band toy model  
\begin{equation}\label{toy-model}
	H = \epsilon_{\vec k}+ (\bex(\vecr) + \gso(\vec k))\cdot \boldsymbol\sigma =\epsilon_{\vec k} + \vec n(\vecr,\vec k)\cdot\boldsymbol\sigma
\end{equation}
where $\boldsymbol \sigma$ is the vector of Pauli matrices, $\bex(\vecr)$ the exchange field arising
from the magnetic texture, $\gso(\vec k)$ the SOI field, and $\vec n=\bex+\gso$.
From Eqs.~(\ref{berry-connection}), (\ref{berry-curvature}) and (\ref{semiclass-eshift}) one finds
\begin{eqnarray}
	\Omega_{\pm,ij} &=& \mp\tfrac12 \hatn \cdot \left(\tfrac{\partial}{\partial x_i} \hatn \times \tfrac{\partial}{\partial x_j} \hatn\right) \label{toymodel-berrycurv}\\
	\delta\epsilon_+ &=& \delta\epsilon_- = |\vec n|\,\textstyle{\sum}_{i=1}^3 \Omega_{+,ii}^{\text{Rk}}. \label{toymodel-eshift}
\end{eqnarray}
where $\pm$ labels the minority and majority band, respectively.
To analyze the model Eq.~(\ref{toy-model}) analytically, we consider the limit of weak SOI parameter $\lambda_{\rm so}$ with $|\gso|/|\bex|\sim \lambda_{\rm so}$.
SOI also controls the size of skyrmions as their formation is driven by DM interactions.
For skyrmion lattices in chiral magnets, the diameter of the skyrmions is proportional to $1/\lambda_{\rm so}$ \cite{muhlbauer_skyrmion_2009}.
Expanding Eqs.~(\ref{toymodel-berrycurv}) and (\ref{toymodel-eshift}) in $\lambdaso$ shows that both $\delta\epsilon_\pm$ and {\em all} components of $\Omega$ are of order $\lambda_{\rm so}^2$.
This argument shows that the expansion in powers of $\Omega$ used in the derivation of Eq.~(\ref{semiclass-deltaf}) is valid for chiral magnets with weak SOI.
Remarkably, all factors of $\lambdaso$ cancel, when the total charge of a single skyrmion $\delta Q^{(1)}=\int \delta \rho^{(1)} \,d^2\!R$ is calculated using Eq.~(\ref{rho1}).
Assuming that the exchange field $\bex$ is small compared to the Fermi energy, we find 
\begin{align}
D_{ij} &\approx   \frac{|\bex|^2}{3}  \int\!\frac{d^2k}{(2 \pi)^2} \; \frac{\partial g^{\rm so}_i}{\partial k_j} \;f''(\epsilon_{\vec k}) \\
	\delta  Q^{(1)} &\approx  \frac{2}{3} e \int\!\frac{d^2R\, d^2k}{(2 \pi)^2}|\bex|^3 \; \Omega^{\text{Rk}}_{+,ii} \;f'''(\epsilon_{\vec k}) \nonumber
\end{align}
To obtain a qualitative estimate, we assume $|\partial g^{\rm so}/\partial k|\sim \lambdaso E_F a$ where $E_F$ is the Fermi energy and $a$ the lattice constant. According to Neutron scattering experiments \cite{muhlbauer_skyrmion_2009,adams_long_range_2011}, the skyrmion lattice in MnSi is well described by
\begin{eqnarray}\label{skyrmion}
	\vec \bex(\vecr) = B_0 \hat{\vec z} + B_1 \sum_{n=0}^2 &[&(\hat{\vec z}\times\hat{\vht{\xi}}_n) \sin(q_0\, \hat{\vht{\xi}}_n\cdot \vec R)\notag\\
	&& \quad + \hat{\vec z} \cos(q_0\, \hat{\vht{\xi}}_n\cdot \vec R)],
\end{eqnarray}
where $q_0\approx 2 \pi/190 {\rm \AA}$, $\hat{\vec z}=(0,0,1)$ is the unit vector parallel to a small magnetic field stabilizing the skyrmion 
lattice, $\hat{\vht{\xi}}_n = (\cos(2\pi n/3),\sin(2\pi n/3),0)$.
From mean-field calculations \cite{adams_long_range_2011} $B_1/B_0\approx-1.5$ is obtained.
As $q_0$ is linear in $\lambdaso$, we set $q_0 =\lambdaso 2 \pi /a$. In this model, we obtain
\begin{eqnarray}
D_{ij} \sim \lambdaso \, \delta_{ij} \,  \frac{E_F}{a} \frac{B_0^2}{E_F^2},
 \qquad \delta Q^{(1)} \sim  e   \frac{B_0^2}{E_F^2}. 
\end{eqnarray}
As expected, the DM interaction is linear in SOI and quadratic in the magnetization. Interestingly,
the skyrmion charge is {\em independent} of the SOI strength (when screening is ignored) but proportional to the square of the local magnetization. 
These main conclusions remain valid when we calculate the charge with {\em ab initio}  methods using the real band structure of a complex material, see below. 

Based on the electronic structure of MnSi obtained within LDA 
we compute $D_{ij}$ and $G_{ij}$ at $T=0$
using Wannier functions~\cite{wannier90code,WannierPaper} to reduce the computational burden 
(see~\cite{supplement} for computational details). Furthermore, we approximate $V$, $\bm E$ and $\bm B$ in Eq.~(\ref{eq_local_hamiltonian}) by their value for vanishing spin-orbit coupling. This allows to 
perform the calculation at $\bm B^{\rm ext}=0$ using that  $\bm B\| \hatn$. The torque is then simply given by $\bm T=\bm M\times \bm B$.
For a left-handed crystal structure we obtain  $D_{ij}=-D\delta_{ij}$ with 
\begin{equation} D=-4.1 \text{ meV\r{A}} \text{\ per 8 atom cell}.\end{equation}
An experimental value for $D$ can be obtained from Neutron scattering in the helical phase of MnSi, because  a finite $D$
shifts the minimum of $E(q)=Dq+Jq^2$ from $q=0$ to $q=-D/(2J)$ for a left-handed spiral. Using $J=52$ meV\r{A}$^2$ per 8 atom cell~\cite{magnetic_excitations_ishikawa}
and $q=2\pi/190{\rm \AA}$ leads to an experimental value of $D=-3.43$ meV\r{A} in good agreement with our result.

\begin{figure}[t]
\centering
\includegraphics[width=8.5cm]{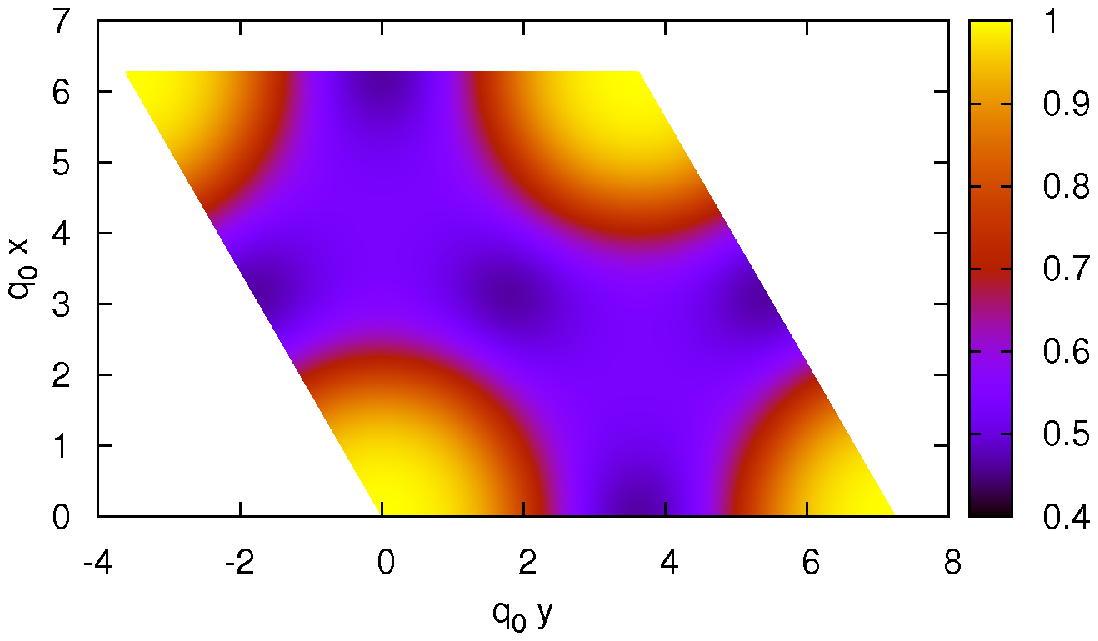}
\caption{\label{fig_skyrmion_charge_density}
Normalized free energy density $\delta F^{(1)}(\vecr)/\delta F^{(1)}(0)$ and 
normalized charge density $\delta \rho^{(1)}(\vecr)/\delta \rho^{(1)}(0)$  within
the magnetic unit cell. The minimal free energy density is given 
by $\delta F^{(1)}(0)=$-0.0018meV/\r{A}$^3$, the total free energy is reduced by 231meV per skyrmion
and layer.
The maximal charge density amounts 
to $\delta \rho^{(1)}(0)=1.95\cdot 10^{-6}e/\mathring{{\rm A}}^3$, 
the total charge per skyrmion and layer 
is 0.246$e$. 
The skyrmion center is located at the origin, as in Eq.~(\ref{skyrmion}).
}
\end{figure}

Next, we discuss the manifestations of phase space Berry phases on the skyrmions in MnSi. As $G_{ij}\propto D_{ij} \propto \delta_{ij}$  by symmetry,  $\delta F^{(1)}(\vecr)$
and  $\delta \rho^{(1)}(\vecr)$ are proportional to each other and can therefore be shown in a single plot,
see Fig.~\ref{fig_skyrmion_charge_density}, where we used
$\hatn(\vecr)=\vec b^{\rm ex}/|\vec b^{\rm ex}|$ (with $\vec b^{\rm ex}$ from Eq.~(\ref{skyrmion})).
Integrating $\delta F^{(1)}(\vecr)$ over the magnetic unit cell
we obtain a free energy reduction of 231 meV. 
Both charge density and free energy density are maximal in the center of the skyrmion located at $(0,0)$. Integrating $\delta \rho^{(1)}(\vecr)$ over the magnetic unit cell
we obtain the charge of 0.246$e$. However, $\delta \rho^{(1)}(\vecr)$ is strongly screened due to the short 
$\lambda_{TF}=\sqrt{\epsilon_0/(e^2 N_F)}\approx 0.224 \mathring{{\rm A}}$, where 
$N_F\approx0.11/(e{\rm V}\mathring{{\rm A}}^3)$ is the density of states at the 
Fermi level obtained in our LDA calculations.
The resulting screened charge density
varies between $\rho_{\rm tot}^{\rm max}\approx 4.5\cdot 10^{-11}e/\mathring{{\rm A}}^3$ 
close to the core and $\rho_{\rm tot}^{\rm min}\approx -4.1\cdot 10^{-11}e/\mathring{{\rm A}}^3$ 
between two skyrmions (see Ref.~\cite{supplement} for illustration).

Our analysis has shown that mixed real-space/momentum-space Berry phases are quantitatively important in materials like MnSi. Energetically, they are the driving force for the formation of magnetic textures and lead to a redistribution of charge in the skyrmion phase which we calculated using {\it ab initio} methods.
For the future, it will be interesting to investigate how the phase space Berry curvature $\Omega^{\rm Rk}$
affects the Hall effect. As in MnSi the contributions arising from the topological Hall effect, i.e., from $\Omega^{\rm RR}$, and the anomalous Hall effect 
due to  $\Omega^{\rm kk}$, are of similar magnitude, we also expect substantial contributions from $\Omega^{\rm Rk}$.

\begin{acknowledgments}
We thank S. Bl\"ugel, H. Geiges, C. Pfleiderer, M. Zirnbauer and, especially, A. Altland for illuminating discussions.
Financial support of the DFG (SFB TR 12, FOR 960), funding under the HGF-YIG programme VH-NG-513 and from Deutsche Telekom Stiftung (R.B.) and computing time on the supercomputers \mbox{JUQUEEN} and \mbox{JUROPA} at J\"ulich Supercomputing Center are gratefully acknowledged.
\end{acknowledgments}

\bibliography{letter}

\end{document}


\renewcommand{\theequation}{S\arabic{equation}} 
\renewcommand{\thepage}{S\arabic{page}} 
\renewcommand{\thesection}{S\arabic{section}}  
\renewcommand{\thetable}{S\arabic{table}}  
\renewcommand{\thefigure}{S\arabic{figure}}

\title{Supplemental material for ``Phase-Space Berry phases in Chiral Magnets: Dzyaloshinskii-Moriya Interaction and the Charge of Skyrmions''}
\author{Frank Freimuth$^{1}$}
\email{f.freimuth@fz-juelich.de}
\author{Robert Bamler$^{2}$}
\author{Yuriy Mokrousov$^{1}$}
\author{Achim Rosch$^{2}$}
\affiliation{$^1$Peter Gr\"unberg Institut and Institute for Advanced Simulation,
Forschungszentrum J\"ulich and JARA, 52425 J\"ulich, Germany}
\affiliation{$^2$Institute for Theoretical Physics, Universit\"at zu K\"oln, D-50937 K\"oln, Germany}
\date{\today}
\begin{abstract}
In this supplemental material we give the computational details of the 
density functional theory calculations, provide a coordinate-independent discussion of the phase space volume,
derive the expressions for the Berry-phase correction to the density of states and the energy shift from a quantum-mechanical gradient expansion
and discuss the charge quantization for skyrmions  in insulators.
\end{abstract}

\maketitle

\section*{Computational details of the \textit{ab initio} calculations}
From the full-potential linearized augmented-plane-wave code {\tt FLEUR}~\cite{fleurcode} 
the electronic structure 
of MnSi was obtained within the local 
density approximation~\cite{lda_mjw} to density functional theory. 
The atomic coordinates and lattice parameter ($a$=4.558\AA) of the 8 atom unit cell of 
MnSi as given in~\cite{MnSi_Pickett}, 
muffin-tin radii of 2.12$a_0$ for both Mn and Si, and a plane-wave cutoff of 3.7$a_0^{-1}$ were
used in the calculations ($a_0=0.529177$\AA\, is Bohr's radius). The basis set was supplemented with
local orbitals for the Mn 3s and 3p states. The unconstrained spin moment per
formula unit is 0.94$\mu_{\rm B}$ and thus larger than the measured spin moment by more than a factor of 2.
We constrained the spin moment per formula unit to the value of 0.4$\mu_{\rm B}$.
From the relativistic first-principles Bloch functions of 100
bands given on an
8x8x8 $\vn{k}$ mesh we constructed 64 
relativistic maximally localized Wannier functions 
using disentanglement within the Wannier90 code~\cite{wannier90}.  
The lowest 40 bands in the valence window are the 32 local orbitals plus 8 
Mn 4s bands. These were skipped, i.e., the 100 bands from which the Wannier functions were disentangled are
bands 41 to 140.

Based on Wannier interpolation~\cite{wannier_interpolation,wannier_rmp} we 
evaluated $D_{ij}(\hat{\vn{n}})$ for the 001, 111, and 110
directions $\hat{\vn{n}}$ of magnetization using a 512x512x512 interpolation mesh.
We find that to very good approximation
\bege\label{eq_dmi}
\vn{D}_{j}(\hat{\vn{n}})=
D_{ij}(\hat{\vn{n}})\hat{\vn{e}}_{i}=
D\hat{\vn{n}}\times(\hat{\vn{n}}\times \hat{\vn{e}}_{j}),
\ee
where $\hat{\vn{e}}_{1}=\hat{\vn{x}}$,
$\hat{\vn{e}}_{2}=\hat{\vn{y}}$ 
and $\hat{\vn{e}}_{3}=\hat{\vn{z}}$ are unit vectors of the cartesian coordinate system and a single
parameter $D=-7.69a_{0}/V$meV describes the amplitude of DMI, with $V=a^3$ the volume of the unit cell.
Eq.~\eqref{eq_dmi} neglects the anisotropy of $\vn{D}_{j}(\hat{\vn{n}})$, which is small according
to our calculations. 

Using Eq.~\eqref{eq_dmi} we can express $\delta F^{(1)}(\vecr)$ as follows:
\bege
\begin{aligned}
\delta F^{(1)}(\vecr)&=\vn{D}_{i}(\hat{\vn{n}})
\cdot\left(
\hat{\vec n}
\times 
\frac{\partial \hat{\vec n}}{\partial  R_i} 
\right) 
\\
&=D
\left[
\hat{\vn{n}}\times(\hat{\vn{n}}\times \hat{\vn{e}}_{i})
\right]\cdot
\left[
\hat{\vec n}
\times 
\frac{\partial \hat{\vec n}}{\partial  R_i}
\right]\\
&=D
\frac{\partial \hat{\vec n}}{\partial  R_i}
\cdot
\left[
\left[
\hat{\vn{n}}\times(\hat{\vn{n}}\times \hat{\vn{e}}_{i})
\right]\times \hat{\vn{n}}
\right]\\
&=D
\frac{\partial \hat{\vec n}}{\partial  R_i}
\cdot
\left[
\hat{\vn{n}}\times \hat{\vn{e}}_{i}
\right]\\
&=D
\left[
\hat{\vn{n}}
\times
\vn{\nabla}
\right]
\cdot
\hat{\vn{n}}
\\
&=D
\hat{\vn{n}}\cdot
\left[
\vn{\nabla}
\times 
\hat{\vn{n}}
\right].
\end{aligned}
\ee

In the skyrmion lattice in MnSi, only the 
derivatives $\partial_{x}\hat{\vn{n}}$ 
and $\partial_y \hat{\vn{n}}$ contribute
and $\partial_{z} \hat{\vn{n}}=0$. Thus, we have
\bege
\vn{\nabla}
\times 
\hat{\vn{n}}=
\left(
\begin{array}{c}
-\sin\theta\,\frac{\partial \theta}{\partial y}\\
\phantom{\sin\theta}\\
\sin\theta\,\frac{\partial \theta}{\partial x}\\
\phantom{\sin\theta}\\
\cos\theta\sin\phi\frac{\partial \theta}{\partial x}+
\sin\theta\cos\phi\frac{\partial \phi}{\partial x}-\\
\cos\theta\cos\phi\frac{\partial \theta}{\partial y}+
\sin\theta\sin\phi\frac{\partial \phi}{\partial y}\phantom{\frac{\frac{T}{T}}{T}}
\end{array}
\right),
\ee
yielding an alternative expression for $\delta F^{(1)}(\vecr)$ in terms of
the azimuthal and polar angles of the exchange field and their derivatives:
\bege
\begin{aligned}
\delta F^{(1)}&(\vecr)=
D\Bigg[
\sin\phi\frac{\partial \theta}{\partial x}
-\cos\phi\frac{\partial \theta}{\partial y}+
\\
&+\sin\theta\cos\theta
\bigg(
\!\!\!\cos\phi\frac{\partial\phi}{\partial x}
+
\sin\phi\frac{\partial\phi}{\partial y}
\bigg)
\Bigg]\\
&=
Dq_0\mathcal{Q}(q_0 x,q_0 y),
\end{aligned}
\ee
where
\bege
\begin{aligned}
\mathcal{Q}(q_0 x,q_0 y)=\Bigg[
\sin\phi\frac{\partial \theta}{\partial(q_0 x)}
-\cos\phi\frac{\partial \theta}{\partial(q_0 y)}+&
\\
+\sin\theta\cos\theta
\bigg(
\!\!\!\cos\phi\frac{\partial\phi}{\partial(q_0 x)}
+
\sin\phi\frac{\partial\phi}{\partial(q_0 y)}
\bigg)
\Bigg].&\\
\end{aligned}
\ee

We define the free energy per skyrmion $\delta E^{(1)}$  as integral of $\delta F^{(1)}(\vecr)$ over the magnetic unit cell,
where we set the extension of the magnetic cell in $z$ direction equal to the lattice parameter $a$ of the
8 atom unit cell of MnSi.
We obtain
\bege
\begin{aligned}
\delta E^{(1)}=&
a\int \delta F^{(1)}(\vecr)d\,xd\,y\\
=&\frac{aD}{q_0}\int d\,(q_0 x)d\,(q_0 y)\mathcal{Q}(q_0 x,q_0 y)\\
=&39\frac{aD}{q_0}=-39\frac{a\lambda}{2\pi}7.69\frac{a_0}{a^3}{\rm meV}
=-47.7\frac{a_0\lambda}{a^2}{\rm meV}=\\
=&-47.4\frac{190\cdot 0.529177}{(4.558)^2}{\rm meV}=-231{\rm meV}.
\end{aligned}
\ee

Determining the tensor
\bege
t_{ij}=e\int\frac{d^3 k}{(2\pi)^3}\sum_{n}f_{\vn{k}n}B_{\vn{k}nij}
\ee
from Wannier interpolation we get
\bege
t_{ij}(\hat{\vn{n}})=t\hat{\vn{e}}_{i}\cdot[\hat{\vn{n}}\times(\hat{\vn{n}}\times\hat{\vn{e}}_{j})],
\ee 
with t=$-0.091ea_{0}/V$, where small anisotropies of $t_{ij}$ have been neglected.
$t_{ij}$ describes the intrinsic component of the SOI-mediated spin torque per volume to an 
applied electric field in the homogeneous system~\cite{ibcsoit}.
One contribution to $\delta\rho^{(1)}(\vn{R})$ is given by 
\bege
-t_{ij}\hat{\vn{e}}_{i}\cdot
\left(
\hat{\vn{n}}\times\frac{\partial \hat{\vn{n}}}{\partial R_{j}}
\right).
\ee
However, due to the additional Fermi surface term, the complete expression for $\delta\rho^{(1)}(\vn{R})$ is given by 
\bege
\delta\rho^{(1)}(\vn{R}) =eG_{ij}\hat{\vn{e}}_{i}\cdot
\left(
\hat{\vn{n}}\times\frac{\partial \hat{\vn{n}}}{\partial R_{j}}
\right),
\ee
where according to our calculations
\bege
eG_{ij}=g\hat{\vn{e}}_{i}\cdot[\hat{\vn{n}}\times(\hat{\vn{n}}\times\hat{\vn{e}}_{j})],
\ee
with $g=0.0082\frac{ea_{0}}{V}$,
neglecting again the small anisotropies. Similar to rewriting the free energy density above, we obtain
\bege
\delta\rho^{(1)}(\vn{R})=g\hat{\vn{n}}\cdot[\nabla \times \hat{\vn{n}} ]=gq_0\mathcal{Q}(q_0 x,q_0 y).
\ee

We define the charge per skyrmion $\delta Q^{(1)}$  as integral of $\delta \rho^{(1)}(\vecr)$ over the magnetic unit cell,
where we set the extension of the magnetic cell in $z$ direction equal to the lattice parameter $a$ of the
8 atom unit cell of MnSi.
This yields
\bege
\begin{aligned}
\delta Q^{(1)}=&
a\int \delta \rho^{(1)}(\vecr)d\,xd\,y\\
=&\frac{ag}{q_0}\int d\,(q_0 x)d\,(q_0 y)\mathcal{Q}(q_0 x,q_0 y)\\
=&39\frac{ag}{q_0}=39\frac{a\lambda}{2\pi}\cdot 0.0082\frac{ea_0}{a^3}
=0.0509\frac{a_0\lambda}{a^2}e=\\
=&0.0509\frac{190\cdot 0.529}{(4.558)^2}e=0.246e.
\end{aligned}
\ee

\begin{figure}[t]
\centering
\includegraphics[width=8.5cm]{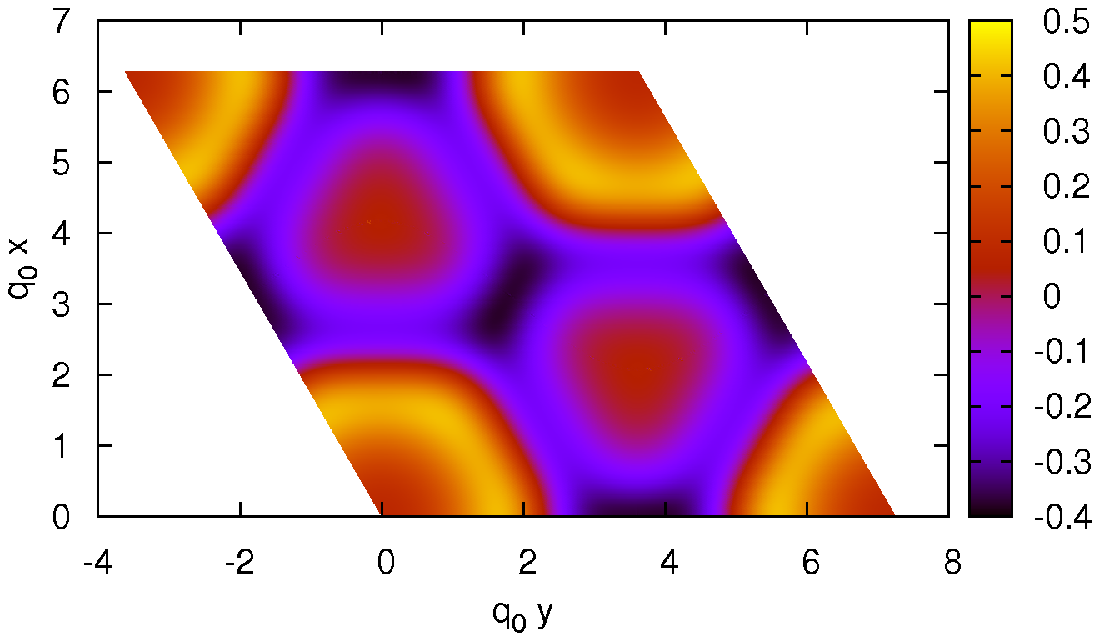}
\caption{\label{fig_skyrmion_charge_density_screened}
Screened charge density $\rho_{\rm tot}/(\delta \rho^{(1)}(0)\lambda^2_{\rm TF}q_0^2)$  within
the magnetic unit cell, with $\delta \rho^{(1)}(0)\lambda^2_{\rm TF}q_0^2=1.07\cdot 10^{-10}e/\mathring{{\rm A}}^3 $.
}
\end{figure}

However, the charge density is strongly screened, see Fig.~\ref{fig_skyrmion_charge_density_screened}.

\section*{Coordinate independent formulation of the phase space volume}

In this section we review the derivation of the volume element, Eq.~(4)
of the main text following mostly the review by Morrison \cite{morrison_hamiltonian_1998} and rewrite some of our formulas using differential forms. This makes the derivation transparent and is manifestly independent of the chosen
coordinate system \cite{warner_foundations_1971,gockeler_differential_1989}. 

Consider a transformation from canonical coordinates $X_i$ with standard Poisson brackets
$\{f,g\}=\frac{\partial f}{\partial X_i} J_{ij} \frac{\partial g}{\partial X_j}$ to a new set of coordinates $x_i=x_i(\vec X)$. The Poission brackets of the (non-canonical) coordinates $x_i$ are given by
\begin{equation}\label{poisson}
\{x_i,x_j\}=\frac{\partial x_i}{\partial X_{i'}} J_{i' j'} \frac{\partial x_j}{\partial X_{j'}}=(\omega^{-1})_{ij}.\end{equation} The natural volume element of the 2d-dimensional phase space is obtained from the Jacobi determinant $\left| \partial X_i/\partial x_j \right|$.
\begin{equation} \label{space1}
d V = \frac{d^{2 d} X}{(2 \pi)^d} = \left| \frac{\partial X_i}{\partial x_j} \right| \frac{d^{2 d} x}{(2 \pi)^d}=\sqrt{\det \omega} \frac{d^{2 d} x}{(2 \pi)^d}
\end{equation}
where we used that $\det \omega=(\left| \partial x_i/\partial X_j \right|^2 \det J)^{-1}=\left| \partial X_i/\partial x_j \right|^2$ as $\det J=1$ and $\det M^{-1}=1/\det M$. 

It is useful to rewrite Eq.~(\ref{space1}) using  that
 the phase space volume $dV$ is independent of the coordinate system.
 In canonical coordinates
we define the  2-form $\hat \omega$ from the inverse of  $J^{-1}=-J$ using
\begin{eqnarray}
\hat{\omega}&=&\frac{1}{2} (J^{-1})_{ij} \, d X^i \wedge dX^j \label{omega}\\
&=&\frac{1}{2} \frac{\partial X_{i'}}{\partial x_i} (J^{-1})_{i'j'}  \frac{\partial X_{j'}}{\partial x_j} \, d x^i \wedge dx^j \nonumber \\
&=& \frac{1}{2}\omega_{ij} \, d x^i \wedge d x^j \label{omega2}.
\end{eqnarray}
where we used the definition of $\omega$ from Eq.~(\ref{poisson}). As in  Eq.~(\ref{omega})
$\hat \omega$ is expressed in canonical coordinates, the phase space volume is directly obtained
from the d-fold wedge product $\hat{\omega}^d=\hat{\omega} \wedge \dots \wedge \hat{\omega}$
\begin{eqnarray}\label{space2}
dV=\frac{\hat{\omega}^d}{d! (2 \pi)^d}\;.
\end{eqnarray}
While  Eq.~(\ref{space1}) and Eq.~(\ref{space2}) are equivalent,  Eq.~(\ref{space2}) is much easier to handle due to the missing square-root.

A remarkable aspect is the close relation of Poisson brackets, phase space volume, Berry connections and Chern classes.
The semiclassical equations of motion for an electron in band $n$ in the presence of phase-space Berry phases are given in the main text as
\begin{equation}
	\dot x_i = ((\Omega_n-J)^{-1})_{ij}\; \frac{\partial\epsilon_n}{\partial x_j} \equiv \{x_i,\epsilon_n\}
\end{equation}
where the Poisson brackets are defined by Eq.~\eqref{poisson} with $\omega = \Omega_n-J$.
From Eq.~\eqref{omega2} follows
\begin{eqnarray}
\hat{\omega}&=& \frac{1}{2}((J^{-1})_{ij}+\Omega_{ij}) \, d x^i \wedge d x^j =\hat \omega_0+\hat \Omega \label{omega3}.
\end{eqnarray}
where $\hat{\omega}_0=\frac{1}{2} (J^{-1})_{ij}\, dx^i \wedge dx^j$ is the `canonical' 2-form which 
obtains a correction from the abelian Berry curvature 
\begin{equation}\label{berry}
{\hat{\Omega}}=d \hat{A}=\frac{1}{2}  \Omega_{ij}\,  d x^i \wedge d x^j
\end{equation}
where  $\hat{A}= A_i \, d x_i$ and we have omitted all band indices.
From Eqs.~(\ref{space2}) and (\ref{omega3}), one obtains  Eq.~(4) of the main text. 

The Berry curvature directly gives the first Chern form
\begin{equation}\label{chern}
\hat{c}_1=\frac{\hat \Omega}{2 \pi}.
\end{equation}
Integrals of the wedge product of $m$ such Chern forms, $\int \hat{c}_1^m=\int \hat{c}_1 \wedge \dots \wedge \hat{c}_1$, over compact $2 m$-dimensional  manifolds without boundary define Chern numbers which are quantized to integers \cite{milnor_characteristic_1974}. 
Such wedge products directly show up when expanding $dV$ in powers of $\hat \Omega$ using Eqs. (\ref{space2}) and (\ref{omega3}),
\begin{eqnarray}
dV=\sum_{m=0}^d
\frac{1}{m! (d-m)!} \hat{c}_1^m \wedge \frac{\hat{\omega}_0^{d-m}}{(2 \pi)^{d-m}}\;.
\end{eqnarray}

\section*{Gradient expansion of the Green's function}

In this section we show that the results from the semiclassical derivation, Eqs.~(3) and (4) of the main text, can be reproduced by a gradient expansion of the quantum-mechanical problem.
We follow the derivation in \cite{rammer_quantum_1986}.
A similar gradient expansion was used by Yang et al.~in \cite{yang_skyrmion_2011} for insulators.
For a spatially inhomogeneous system with Green's function $G(\omega; \vec r_1, \vec r_2)$, we introduce the Wigner transform of the Green's function,
\begin{equation}
	\tilde G(\omega; \vec x) \equiv \tilde G(\omega; \vec R, \vec k) = \frac{1}{V}\!\!\int\!\! d^3r\, e^{-i\vec k\cdot\vec r}G(\omega; \vec R\!+\!\frac{\vec r}{2}, \vec R\!-\!\frac{\vec r}{2})
\end{equation}
where $V$ is the volume and $\vec R=\frac12(\vec r_1+\vec r_2)$ and $\vec r=\vec r_1-\vec r_2$ are the center-of mass and relative coordinates, respectively.
The Wigner transform $\tilde K$ of the inverse of the Green's function, $K(\omega; \vec r_1, \vec r_2)\equiv G^{-1}(\omega; \vec r_1, \vec r_2)$, is defined analogously and satisfies the relation
\begin{equation}\label{moyal-convolution}
	e^{\frac{i}{2} J_{ij} \partial^K_i \partial^G_j} \tilde K(\omega; \vec x) \tilde G(\omega; \vec x) = \mathbb 1
\end{equation}
where the derivative $\partial^K_i$ ($\partial^G_j$) acts on $\tilde K$ ($\tilde G$) only and $\mathbb1$ is the unit matrix in band space.
For a smooth spatial variation, expanding the exponential in Eq.~(\ref{moyal-convolution}) leads to $\tilde G \approx \tilde G_0 + \tilde G_1 + \mathcal O(\partial^4)$ where $\tilde G_0(\omega;\vec x)\equiv \tilde K^{-1}(\omega;\vec x)$ is the semi-classical Green's function and
\begin{equation}\label{moyal-greensfct}
	\tilde G_1 = \frac{i}{2}J_{ij} \tilde G_0 (\partial_i \tilde G_0^{-1}) \tilde G_0 (\partial_j \tilde G_0^{-1}) \tilde G_0.
\end{equation}
For a non-interacting system, $\tilde G_0(\omega;\vec x)=(\hbar\omega-H(\vec x))^{-1}$, where $H(\vec x)$ is the semi-classical Hamiltonian.

The charge density is obtained from
\begin{equation}\label{chargedens-summation}
	\rho(\vec R) = eT\sum_{\omega_n} \int\!\frac{d^3k}{(2\pi)^3}\; \text{Tr}[\tilde G(i\omega_n;\vec R,\vec k)]
\end{equation}
where $e$ is the electron charge, $T$ the temperature and $\hbar\omega_n=k_BT\pi (2n+1)$.
From the evaluation of the frequency summation in Eq.~(\ref{chargedens-summation}) we get to first order in spatial gradients (cf. Eq.~(11) of the main text)
\begin{equation}\label{moyal-rho}
	\rho(\vec R) \!=\! e\sum_n\! \int\!\!\frac{d^3k}{(2\pi)^3} \!\left[
		 \frac{\partial f(\epsilon_n(\vec x))}{\partial \epsilon_n(\vec x)} \delta\epsilon_n(\vec x) + f(\epsilon_n(\vec x)) W_n(\vec x)
	\right]
\end{equation}
where $f$ is the Fermi function and
\begin{alignat}{1}
	\delta\epsilon_n(\vec x) &= -\sum_{m\neq n} \text{Im}\left[
		\frac{\langle \vec x n |\frac{\partial H}{\partial R_i}|\vec x m\rangle\langle \vec x m| \frac{\partial H}{\partial k_i}|\vec x n\rangle}{\epsilon_n(\vec x)-\epsilon_m(\vec x)}
	\right] \label{moyal-deltae} \\
	 W_n(\vec x) &= 1+2\sum_{m\neq n}\text{Im}\left[
		\frac{\langle \vec x n |\frac{\partial H}{\partial R_i}|\vec x m\rangle\langle \vec x m| \frac{\partial H}{\partial k_i}|\vec x n\rangle}{(\epsilon_n(\vec x)-\epsilon_m(\vec x))^2}
	\right] = \nonumber\\
	&=1-\sum_{i=1}^3 \Omega_{n,ii}^{\text{Rk}}. \label{moyal-dos}
\end{alignat}
Eq.~(\ref{moyal-deltae}) is equivalent to the semiclassical energy shift, Eq.~(3) of the main text, and Eq.~(\ref{moyal-dos}) is equivalent to Eq.~(4) of the main text up to linear order in $\Omega_n$.
The Free energy can be calculated in a similar way by expanding $F=-T\sum_{\omega_n}\!\text{Tr}\log[-T\tilde G]$ in powers of the gradients.
The result confirms Eq.~(6) of the main text.

\section*{Quantized skyrmion charge in insulators}

In this section, we show that the Berry curvature contribution to the charge in a two-dimensional insulator with abelian Berry curvature is given by the product of the quantized Hall conductivity $\sigma_{xy}$ and the skyrmion number $\Phi_0$.
An abelian Berry curvature arises, e.g., if only a single band is occupied.
We then derive an expression for the charge per length of a skyrmion line in a three-dimensional insulator.

For an abelian Berry curvature in a two-dimensional system, Eq.~(13) of the main text reduces to
\begin{alignat}{1}
	\delta Q &=- \frac{e\,\epsilon_{i j k l}}{8}\int \frac{ d^4 x}{(2 \pi)^2} \Omega_{ij} \Omega_{kl} \nonumber \\
	&= -\frac{e\,\epsilon_{i j k l}}{4} \left(\int \frac{ d x_j\,dx_k\,dx_l}{(2 \pi)^2} A_j\Omega_{kl}\right)_{\!x_i=-\infty}^{\!x_i=+\infty} \label{2dcharge1}
\end{alignat}
where the symbols $\pm\infty$ denote either positions far away from the skyrmion or the boundaries of the Brillouin zone for a space or momentum direction $x_i$, respectively. In the second equality of Eq.~(\ref{2dcharge1}), we used the relation
\begin{alignat}{1}
	\frac{\epsilon_{ijkl}}{8} \Omega_{ij}\Omega_{kl} &= \frac{\epsilon_{ijkl}}{2} \frac{\partial A_j}{\partial x_i} \frac{\partial A_l}{\partial x_k} = \frac{\epsilon_{ijkl}}{2} \frac{\partial}{\partial x_i} \left(A_j\frac{\partial A_l}{\partial x_k}\right) \nonumber \\
	&= \frac{\epsilon_{ijkl}}{4} \frac{\partial}{\partial x_i} \left(A_j\Omega_{kl}\right) \;.
\end{alignat}

In Eq.~(\ref{2dcharge1}), $\Omega_{kl}$ only enters at the boundary of the $x_i$ coordinate.
At the boundary in spatial direction, the magnetization is collinear and therefore $\Omega^{RR}=0=\Omega^{Rk}$.
Thus, if $x_i$ is a spatial coordinate, only terms of the form $A^R_j\Omega^{kk}_{kl}$ contribute to the integral kernel in Eq.~(\ref{2dcharge1}).
If $x_i$ is a momentum coordinate, $\Omega_{kl}$ is evaluated at the boundary of the Brillouin zone.
In an insulator, the charge must be quantized and we can adiabatically deform the Bloch functions such that they are independent of momentum in a narrow stripe around the Brillouin zone boundary.
This is always possible since, in the absence of further symmetries, all non-interacting Hamiltonians of one-dimensional insulators are adiabatically connected \citep{ryu_topological_2010}.
Therefore, only terms of the form $A^k_j\Omega^{RR}_{kl}$ contribute if $x_i$ is a momentum coordinate.
In total, Eq.~(\ref{2dcharge1}) can be written as $\delta Q = \delta Q^R + \delta Q^k$ where
\begin{equation}\label{deltaqr}
	\delta Q^R = -\frac{e\,\epsilon_{ij}}{2} \left(
	\int \frac{dR_j\,d^2 k}{(2 \pi)^2} A^R_j\Omega^{kk}_{xy}
	\right)_{\!R_i=-\infty}^{\!R_i=+\infty}
\end{equation}
and $\delta Q^k$ is defined by formally exchanging all $R$ and $k$.
As the Berry curvature $\Omega^{kk}_{xy}$ in Eq.~(\ref{deltaqr}) is gauge independent, it cannot depend on $R_j$ for a collinear magnetization at $R_i=\pm\infty$.
This implies
\begin{alignat}{1}
	\delta Q^R &= -\frac{e\,\epsilon_{ij}}{2}
	\int\!\frac{d^2 k}{(2 \pi)^2} \left(\Omega^{kk}_{xy} \int\!dR_j (A^R_j)_{\!R_i=-\infty}^{\!R_i=+\infty}
	\right) \nonumber \\
	&= -\frac{e}{2}
	\int\!\frac{d^2 k}{(2 \pi)^2} \left(\Omega^{kk}_{xy} \int\!d^2 R\, \epsilon_{ij}\frac{\partial A^R_j}{\partial R_i}
	\right) \nonumber \\
	&= -\frac{e}{2}
	\int\!\frac{d^2 k}{(2 \pi)^2} \left(\Omega^{kk}_{xy} \int\!d^2 R\, \Omega^{RR}_{xy}
	\right) \nonumber \\
	&= -\frac{1}{2} \sigma_{xy}\Phi_0
\end{alignat}
where $\sigma_{xy}$ ($\Phi_0$) is the quantized integral over $\frac{e^2}{\hbar}\Omega^{kk}_{xy}$ ($\frac{\hbar}{e}\Omega^{RR}_{xy}$) as defined in the main text.
An analogous calculation leads to the same value for $\delta Q^k$.
Thus, we conclude that the Berry curvature contribution to the skyrmion charge in a two-dimensional insulator with abelian Berry curvature is given by
\begin{equation}
	\delta Q=-\sigma_{xy}\Phi_0. \label{charge2dquantized}
\end{equation}

In three dimensional systems, skyrmions form line defects.
From Eqs.~(\ref{space2})-(\ref{omega3}), the Berry curvature contribution to the charge in a three dimensional insulator is given by
\begin{equation}\label{charge3dforms}
	\delta Q = \frac{3e}{3!(2\pi)^3} \int \hat\omega_0\wedge\hat\Omega\wedge\hat\Omega
\end{equation}

To reduce the three dimensional insulator to a set of two-dimensional insulators, we introduce dimensionless coordinates $\vec{\tilde x} \equiv (\vec{\tilde R},\vec{\tilde k})$ such that $\vec R=\tilde R_\alpha \vec a_\alpha$ and $\vec k=\tilde k_\alpha \vec g_\alpha$.
Here, the vectors $\vec a_\alpha$ ($\vec g_\alpha$) are direct (reciprocal) lattice vectors, respectively.
The coordinates $\tilde{\vec x}$ are chosen such that momentum space is periodic in the three coordinate directions $\tilde k_\alpha$.
In these coordinates, the Berry curvature and the symplectic form read
\begin{alignat}{1}
	\hat\Omega &= \frac12 \Omega_{ij}\; \frac{\partial x_i}{\partial\tilde x_\alpha} \frac{\partial x_j}{\partial\tilde x_\beta} \; d\tilde x_\alpha\wedge d\tilde x_\beta
	\equiv \frac12 \tilde\Omega_{\alpha\beta}\; d\tilde x_\alpha\wedge d\tilde x_\beta; \nonumber \\
	\hat\omega_0 &= -\frac12 J_{ij}\; \frac{\partial x_i}{\partial\tilde x_\alpha} \frac{\partial x_j}{\partial\tilde x_\beta}\; d\tilde x_\alpha\wedge d\tilde x_\beta
	 = -\pi J_{\alpha\beta}\; d\tilde x_\alpha\wedge d\tilde x_\beta
\end{alignat}
where we used that $\vec a_\alpha\cdot\vec g_\beta=2\pi\delta_{\alpha\beta}$ in the expression for $\hat\omega_0$.
We write Eq.~(\ref{charge3dforms}) in coordinates $\vec{\tilde x}$,
\begin{alignat}{1}
	\delta Q &= -\frac{e\,\epsilon_{\alpha\beta\gamma\delta\mu\nu}}{16(2\pi)^2} \int\! d^6\tilde x \,
	J_{\alpha\beta} \tilde\Omega_{\gamma\delta} \tilde\Omega_{\mu\nu} \nonumber\\
	&= -e\sum_{\alpha=1}^3 \int\! d \tilde R_\alpha \,d \tilde k_\alpha \left( \frac{\epsilon_{\alpha(\alpha+3)\gamma\delta\mu\nu}}{8}
	\int\!\frac{d^4\tilde x}{(2\pi)^2} \,
	\tilde\Omega_{\gamma\delta} \tilde\Omega_{\mu\nu}\right) \label{6dintegral}
\end{alignat}
where the last integral is over the four dimensional subspace of phase space perpendicular to $(\vec 0,\vec a_\alpha)$ and $(\vec g_\alpha,\vec 0)$.
One can use the same arguments as the ones that lead from Eq.~(\ref{2dcharge1}) to Eq.~(\ref{charge2dquantized}) to see that this integral factorizes,
\begin{alignat}{1}
	&\frac{\epsilon_{\alpha(\alpha+3)\gamma\delta\mu\nu}}{8}
	\int\!\frac{d^4 \tilde x}{(2\pi)^2} \,
	\tilde\Omega_{\gamma\delta} \tilde\Omega_{\mu\nu} = \nonumber\\
	= &\left( \frac{\epsilon_{\alpha\gamma\delta}}{2} \int d\tilde R_{\gamma} d\tilde R_{\delta} \;\tilde\Omega^{RR}_{\gamma\delta} \right)
	\left( \frac{\epsilon_{\alpha\mu\nu}}{2} \int \frac{d\tilde k_{\mu} d\tilde k_{\nu}}{(2\pi)^2} \;\tilde\Omega^{kk}_{\mu\nu} \right) \nonumber\\
	\equiv & n^R_\alpha n^k_\alpha \qquad\text{(no sum over $\alpha$)} \label{3dfactorize}
\end{alignat}
where $n^R_\alpha$ ($n^k_\alpha$) is the real-space (momentum-space) winding number in the plane perpendicular to $\vec g_\alpha$ ($\vec a_\alpha$), respectively.
Excluding, for now, the case where the skyrmion line is parallel to a reciprocal lattice vector $\vec g_\beta$, the skyrmion line pierces all three coordinate planes and $n^R_\alpha=1$ for all $\alpha$.
The final result for the charge (Eq.~(\ref{3d-result})) will be independent of all $n^R_\alpha$ with $\alpha\neq\beta$ if the skyrmion line points into the direction of $\vec g_\beta$.

Thus, for fixed $\alpha$, the term in parenthesis in Eq.~(\ref{6dintegral}) is  quantized and therefore independent of $\tilde R_\alpha$ and $\tilde k_\alpha$.
The remaining integral over $\tilde k_\alpha$ equates to a factor of $1$ and the integral over $\tilde R_\alpha$ gives
\begin{equation}\label{3dinsulator-length}
	\int\!d\tilde R_\alpha
	= \frac{\partial \tilde R_\alpha}{\partial R_i}\int\! d R_i 
	= \frac{(\vec g_\alpha)_i}{2\pi} L_i = \frac{\vec{\hat s}\cdot\vec g_\alpha}{2\pi}L
\end{equation}
where the unit vector $\hat{\vec s}$ points along along the skyrmion line and $L_i=L \hat s_i$ is the projection of the length $L$ of the skyrmion line onto the coordinate direction $R_i$.
Combining Eqs.~(\ref{6dintegral})-(\ref{3dinsulator-length}), we arrive at an expression for the charge per length of a skyrmion line in a three dimensional insulator with abelian Berry curvature,
\begin{equation} \label{3d-result}
	\frac{\delta Q}{L} = -e\sum_{\alpha=1}^3 \frac{\vec{\hat s}\cdot\vec g_\alpha}{2\pi}  n^k_\alpha
\end{equation}
where $\vec{\hat s}$ is the direction of the skyrmion line, $\vec g_\alpha$ are reciprocal lattice vectors, and $n^k_\alpha\in\mathbb Z$ is defined in Eq.~(\ref{3dfactorize}).

\bibliography{letter_supp}